\documentclass[manuscript]{aastex}

\slugcomment{Prepared for ApJL}

\shorttitle{Equilibria for Binary Asteroids}
\shortauthors{Jacobson \& Scheeres}

\usepackage{times,amsmath}				
\usepackage{url}                
\usepackage{color}
\usepackage{graphicx}			
\usepackage{hyperref}

\begin{document}

\title{Long-term Stable Equilibria for Synchronous Binary Asteroids}

\author{Seth A. Jacobson\altaffilmark{1} and Daniel J. Scheeres\altaffilmark{2}}
\affil{Department of Astrophysical and Planetary Sciences, University of Colorado at Boulder, Boulder, CO, 80309 USA}
\affil{Department of Aerospace Engineering Sciences, University of Colorado at Boulder, Boulder, CO, 80309 USA}

\begin{abstract}
Synchronous binary asteroids may exist in a long-term stable equilibrium, where the opposing torques from mutual body tides and the binary YORP (BYORP) effect cancel. Interior of this equilibrium, mutual body tides are stronger than the BYORP effect and the mutual orbit semi-major axis expands to the equilibrium; outside of the equilibrium, the BYORP effect dominates the evolution and the system semi-major axis will contract to the equilibrium. If the observed population of small (0.1 - 10 km diameter) synchronous binaries are in static configurations that are no longer evolving, then this would be confirmed by a null result in the observational tests for the BYORP effect. The confirmed existence of this equilibrium combined with a shape model of the secondary of the system enables the direct study of asteroid geophysics through the tidal theory. The observed synchronous asteroid population cannot exist in this equilibrium if described by the canonical ``monolithic'' geophysical model.  The ``rubble pile'' geophysical model proposed by~\citet{Goldreich2009} is sufficient, however it predicts a tidal Love number directly proportional to the radius of the asteroid, while the best fit to the data predicts a tidal Love number inversely proportional to the radius. This deviation from the canonical and ~\citet{Goldreich2009} models motivates future study of asteroid geophysics. Ongoing BYORP detection campaigns will determine whether these systems are in an equilibrium, and future determination of secondary shapes will allow direct determination of asteroid geophysical parameters. 
 \end{abstract}
 
 \keywords{Celestial mechanics --- Minor planets, asteroids: general --- Planets and satellites: dynamical evolution and stability --- Planets and satellites: fundamental parameters --- Planets and satellites: interiors}

\section{Introduction}
\label{sec:Introduction}
Synchronous binary asteroid systems may now provide a method of directly probing an asteroid's internal structure. They may be inhabiting a long-term stable equilibrium created by the opposing torques from mutual body tides and the binary YORP (BYORP) effect. From the tidal theory, this equilibrium would allow direct study of an asteroid's geophysics for the first time. This potential equilibrium was suggested as a possibility by~\citet{Goldreich2009}, but not fully explored.

For this long-term stable equilibrium to exist, the internal structure of binary asteroids must be different than the usually assumed ``monolithic'' canonical internal structure developed for astrophysical bodies~\citep{Darwin1879, Love1927,Goldreich1963, Kaula1964}. Since the canonical theory was determined by studying much larger, hydrostatically balanced bodies under simplifying continuum mechanical approximations, it is not surprising that this theory does not encompass small, very low gravity, ``rubble pile'' asteroids.~\citet{Goldreich2009} derived an alternative theory for the internal structure by modifying the continuum mechanical approximations to include the effect of voids in the ``rubble pile'' structure. As they noted, this new internal structure does allow for this equilibrium to be stable in eccentricity.

We study the internal structure of synchronous binary asteroid systems by making two observationally and theoretically motivated assumptions. First, these binaries formed via rotational fission and so share a common parent body~\citep{Margot2002, Scheeres2007, Walsh2008, Pravec2010, Jacobson2011}. Each component of the binary will share the same intensive properties such as density $\rho$, tidal dissipation number $Q$, rigidity $\mu$, and yield strain $\epsilon_Y$. Extensive properties such as radius $R$, mass ratio\footnote{Defined as the mass of the smaller (secondary) component divided by the mass of the larger (primary) component of the binary.} $q$, and tidal Love number $k$ depend on the absolute sizes of the binary members.

Second, these systems are observed proportional to the total time they spend in each state of their evolution. Binary asteroid systems comprise a significant fraction ($15 \pm 4\%$) of the NEA population~\citep{Margot2002, Pravec2006}. Both~\citet{Cuk2007} and~\citet{McMahon2010b} determined that if the BYORP effect dominates the evolution, then synchronous binaries can disrupt in much less than a million years, requiring frequent binary creation to maintain the observed population. If binaries are trapped in a long-term stable equilibrium, then binary creation could be infrequent. This conclusion is corroborated by evidence that the binary formation process from rotational fission is inefficient, requiring many rotational fission events per asteroid (taking many YORP timescales), in order to create a stable binary system that does not immediately disrupt~\citep{Jacobson2011}. The observed synchronous binary population may be residing in this equilibrium.

After the formation of a stable binary asteroid system via rotational fission (or some other mechanism), tides immediately begin to dissipate energy from the system through the mechanical stressing of each body. The fastest process is the synchronization of the secondary which occurs because of tides raised on the secondary by the primary~\citep{Goldreich1963, Goldreich2009}. This process is often referred to as tidal locking of the satellite. After the secondary has been synchronized, the system will evolve due to both tides and the BYORP effect, which requires a synchronous secondary. 
\subsection{Tidal Evolution}
\label{sec:Tides}
 Relative motion between components in a binary system leads to tidal dissipation of energy and the transfer of angular momentum between spin and orbit states. Assuming spherical, homogenous bodies with identical compositions and a mutual orbit with low eccentricity, an asteroid's first order geophysics can be characterized by two parameters: the tidal Love number and the tidal dissipation number. The tidal Love number $k$ is the ratio of the additional gravitational potential produced by the redistribution of mass relative to the deforming potential, and theoretically is thought to depend on the size and internal properties of the body. The tidal dissipation number $Q$ describes how effective the body is at tidally dissipating energy. It is a quality factor defined as: $ Q = 2 \pi E_0  / \oint \frac{dE}{dt}\ dt $, where $E_0$ is the maximum energy stored in the tidal distortion during a cycle and $\oint \frac{dE}{dt}\ dt$ is the energy dissipated over one cycle. 

As stated before, tides raised on the primary by the secondary synchronize the spin of the secondary to the mutual orbit period; this is the fastest tidal evolutionary process~\citep{Goldreich1963, Goldreich2009}. Tidal evolution continues after synchronization of the secondary. As tides dissipate energy from the rotation states of binary members, the semi-major axis and the eccentricity of the mutual orbit change over time. Only after the rotation period of the primary is synchronous with the mutual orbital period as well and the system is thus doubly (completely) synchronous will binary systems not experience any tidal evolution.

When the system is singly synchronous, tides raised on the primary by the secondary cause the semi-major axis to expand. From the first order theory, the semi-major axis time rate of change due to tides is:
\begin{equation}
\dot{a}_T = 3 \frac{k_p}{Q} \left(\frac{\omega_d}{a^{11/2}}\right) q \sqrt{1+q} 
\end{equation}
\noindent where~$a$ is the semi-major axis measured in primary radii $R_p$, $k_p$ is the tidal Love number of the primary,~$\omega_d = ( 4 \pi G \rho / 3)^{1/2}$ is the surface disruption spin limit for a sphere, and~$G$ is the gravitational constant~\citep{Murray1999}. Due to the fast rotation of the primary compared to the mean motion, tidal evolution always expands the semi-major axis of the binary system.

Tides raised on the primary by the secondary cause the eccentricity to grow, while tides raised on the secondary by the primary cause eccentricity to damp. From the first order theory, the tidal eccentricity time rate of change due to tides is:
\begin{equation}
\dot{e}_T = \frac{57 k_p q^{1/3} - 84 k_s}{8 Q} \left(\frac{\omega_d e}{a^{13/2}}\right) q^{2/3} \sqrt{1+q}
\end{equation}
\noindent where $e$ is the eccentricity and $k_s$ is the tidal Love number of the secondary~\citep{Murray1999}. Tidal evolution can cause excitation or damping of the eccentricity depending on the system mass ratio and the tidal Love numbers of each component.
 
There are two developed theories for the internal structure of asteroids: ``monolith'' and ``rubble pile.'' Evidence that asteroids have a ``rubble pile'' internal structure rather than a ``monolithic'' interior includes: the Hayabusa mission to Itokawa showing no obvious impact craters and the appearance of a structure made entirely from shattered fragments of different size scales~\citep{Fujiwara2006}, numerical modeling of collisions of asteroids~\citep{Michel2001}, mass and volume measurements from the NEAR Shoemaker flyby of Mathilde~\citep{Yeomans1997} and radar observations of 1999 KW$_4$~\citep{Ostro2006} showing mean densities that are lower than their constitutive elements, the rotational speed limit period ($\sim 2.2$ hours) observed amongst asteroids with diameters larger than $\sim200$ m, which corresponds to the critical disruption spin rate of a self-gravitating, strengthless body~\citep{Pravec2007a}, and that rotational fission of strengthless bodies is responsible for the asteroid pair population~\citep{Pravec2010}. All of this evidence suggests that a ``rubble pile'' internal structure is a more realistic assessment than a ``monolith" internal structure, however for completeness we will consider both theories. The two theories become distinct when assessing the functional form of the tidal Love number. 

The dimensionless rigidity of the body $\tilde{\mu}$, which can be thought of as the ratio of the fluid strain to the elastic strain, can be used to defined the tidal Love number: $k = 1.5 / ( 1 + \tilde{\mu} ) \approx 1.5 / \tilde{\mu}$, this approximation is true when the fluid strain dominates the elastic strain~\citep{Goldreich2009}. According to the canonical ``monolith" theory the dimensionless rigidity has the form $\tilde{\mu} =19 \mu / (2 \omega_d^2 \rho R^2)$, where $\mu$ is the rigidity of the body. To first order, the tidal Love number for a ``monolith'' depends on the size of the body as $k_M \propto R^2$.

\citet{Goldreich2009} developed an alternative ``rubble pile" tidal Love number theory by studying how introducing voids increases the stress across contact areas. The dimensionless rigidity of a ``rubble pile'' is smaller than that of a ``monolith'' of the same size. Using conceptual and dimensional arguments,~\citet{Goldreich2009} determine that the ``rubble pile'' dimensionless rigidity should scale as $\tilde{\mu}_R \gtrsim (\tilde{\mu} / \epsilon_Y )^{1/2} $ where $\epsilon_Y$ is the yield strain. To first order, the tidal Love number for a ``rubble pile'' depends on the size of the body as $k_R \propto R$. Both of these theories will be explored in later sections.
\subsection{BYORP Evolution}
\label{sec:BYORP}
The BYORP effect is the summation of radiative effects on a synchronous secondary~\citep{Cuk2005}.~\citet{McMahon2010a} using averaging theory showed that this effect secularly evolves both the semi-major axis and the eccentricity, because radiative forces acting on asymmetries in the shape of the secondary create torques on the mutual orbit that persist after averaging over the mutual orbit, the heliocentric orbit, and the precession of the node. To first order, the evolution of the semi-major axis and the eccentricity only depends upon a single constant term $B$ that represents the averaged acceleration in the direction parallel to the motion of the secondary~\citep{McMahon2010b}. The BYORP coefficient $B$ depends only on the shape of the secondary, and can be thought of as a ratio relating the asymmetric area, on which the average force of the BYORP effect acts, to the total area. A symmetric body--a sphere or an ellipsoid--has a value:~$B = 0$. The BYORP coefficient has a maximum magnitude:~$B = 2$ in either direction, but commonly has a small value: $B \sim 10^{-3}$. The BYORP effect can either expand or shrink the semi-major axis, with the sign of the eccentricity evolution always opposite that of the semi-major axis evolution. From the first order theory, the evolution of the semi-major axis and eccentricity is:
\begin{equation}
\dot{a}_B = \pm \frac{3 H_\odot B}{2 \pi} \left(  \frac{ a^{3/2}}{\omega_d \rho R_p^2} \right)  \frac{\sqrt{1+q}}{q^{1/3}}
\end{equation}
\begin{equation}
\dot{e}_B = \mp \frac{3 H_\odot B}{8 \pi} \left(  \frac{ a^{1/2} e}{\omega_d \rho R_p^2} \right)  \frac{\sqrt{1+q}}{q^{1/3}}
\end{equation}
\noindent where~$H_\odot = F_\odot / ( a_\odot^2 \sqrt{1-e_\odot^2} )$, $F_\odot$ is the solar radiation constant,~and $a_ \odot$ and $e_ \odot$ are the heliocentric semi-major axis and eccentricity~\citep{McMahon2010b}.
\section{Joint Evolution}
\label{sec:Joint}
Synchronous binary asteroids will evolve under the influence of both tides and BYORP. There are two scenarios: joint expansive evolution and joint opposing evolution depending on the direction of the BYORP torque, which nominally has an equal chance of pointing in either direction. Both tides and the BYORP effect can change the energy of the system over time, and the BYORP effect can also changes the system angular momentum. It is convenient to define a parameter $A$, which compares the strengths of these two effects on the semi-major axis of the system, and $E$ which does the same for the eccentricity:
\begin{equation}
A = \frac{|\dot{a}_T|}{|\dot{a}_B|} = \frac{2 \pi \omega_d^2 \rho k_p R_p^2 q^{4/3}}{B H_\odot Q a^7}
\label{eqn:semimajoraxisevolution}
\end{equation}
\begin{equation}
E = \frac{|\dot{e}_T|}{|\dot{e}_B|} = \frac{|19 - 28 K|}{2} A
\end{equation}
\noindent where $K = k_s / (k_p q^{1/3})$ is an important tidal Love number relation. For the ``monolith'' model $k \propto R^2$, so $K = q^{1/3}$, and for the ``rubble pile'' model $k \propto R$, so $K = 1$. If $A < 1$ then the BYORP effect dominates semi-major axis evolution, $A > 1$ then tides dominate the semi-major axis evolution, and if $A = 1$ then the two are balanced and the system semi-major axis will not evolve via either effect. Which effect dominates eccentricity is shown similarly by $E$.
\subsection{Joint Expansive Evolution}
\label{sec:Expansive}
During joint expansive evolution, both effects are growing the semi-major axis. Tides are removing energy by spinning down the primary and transferring angular momentum from the primary spin to the orbit, while the BYORP effect is directly adding both energy and angular momentum to the orbit. The eccentricity of the mutual orbit can be excited or damped depending on the system mass ratio and the tidal Love numbers of the components. The condition for stability or damping in eccentricity is: 
\begin{equation}
\frac{19 - 28 K}{2} A \leq 1
\end{equation}
For ``monoliths,'' $K = q^{1/3}$ and so low mass ratio systems can have growing eccentricity, if $q < (19/28)^3$ and $A > 2 / (19 - 28 q^{1/3})$. For ``rubble pile'' systems, $K = 1$ and thus the condition is always satisfied, and so the eccentricity of binary asteroids with ``rubble pile'' internal structures will always damp. 

Since the semi-major axis growth is always positive, systems that damp eccentricity will grow towards the Hill sphere and eventually disrupt, disregarding the possible evolutionary paths of~\citet{Cuk2010}. Systems that grow in both semi-major axis and eccentricity will do so until the evolutionary equations in the first order theory no longer apply and more complicated dynamics become dominant. 
 \subsection{Joint Opposing Evolution}
\label{sec:Opposing}
During joint opposing evolution, tides are acting to grow the semi-major axis, but the BYORP effect is acting to shrink it. Thus the system evolves towards an equilibrium point, where these two effects balance. An equilibrium exists for the evolution of the semi-major axis at 
\begin{equation}
a^* = \left( \frac{2 \pi k_p \omega_d^2 \rho R_p^2 q^{4/3} }{ B H_\odot Q} \right)^{1/7}
\end{equation}
This is a stable equilibrium, and regardless of the initial semi-major axis, the system will evolve in semi-major axis to this equilibrium point. Tides are still transferring angular momentum to the orbit and removing energy from the system, but now the BYORP effect is removing angular momentum and energy from the system. Interior to the equilibrium semi-major axis, tides dominate and the orbit grows due to the increase of angular momentum to the orbit. Exterior to the equilibrium, the BYORP effect controls the evolution and the orbit shrinks due to the decrease of orbital angular momentum. At the equilibrium the amount of angular momentum removed from the orbit equals the amount tidally transferred into the orbit from the primary, de-spinning it. The torque on the primary is equal in strength to the BYORP torque $\Gamma_B$ and so the rate of the de-spinning of the primary is:
\begin{equation}
\dot{\omega}_{p} = - \frac{\Gamma_B}{I_p} = - \frac{45 H_\odot B}{32 \pi^2} \left( \frac{a_p}{\rho^2 R_p^7} \right) \frac{1+q}{q}
\end{equation}

The eccentricity of the mutual orbit can be excited or damped depending on the system mass ratio and the tidal Love numbers of the components. The condition for stable evolution without growth in eccentricity is:
\begin{equation}
\frac{19 - 28 K}{2} A \leq -1 
\end{equation}
For ``monolith'' asteroids, $K = q^{1/3}$ and so it is possible for any mass ratio system to be unstable in eccentricity if $A < 2 / (28 q^{1/3} - 19)$ and in fact, any system with a mass ratio $q < (19/28)^{3} \approx 0.31$ will be unstable regardless of $A$. At the equilibrium, $A = 1$ and so binaries with ``monolith'' internal structures will grow in eccentricity if the system mass ratio $q < 27/64 \approx 0.42$. This would exclude the observed synchronous binaries from existing in this equilibrium condition, since they all have mass ratios $q < 0.2$ as shown in Table~\ref{tab:BQk}. Therefore these systems might enter equilibrium but could not remain there for a long period of time since their eccentricities would continue to grow.

For binary asteroids with ``rubble pile'' internal structures, $K = 1$ and the stability condition is always satisfied, and thus the eccentricity will always be damped. Thus the observed synchronous binary population can exist in the equilibrium without growth in eccentricity.
\section{Implications for the Synchronous Binary Asteroid Population:}
\label{sec:Implications}
\begin{figure}[h!]
\begin{center}
\includegraphics[width=0.7\columnwidth]{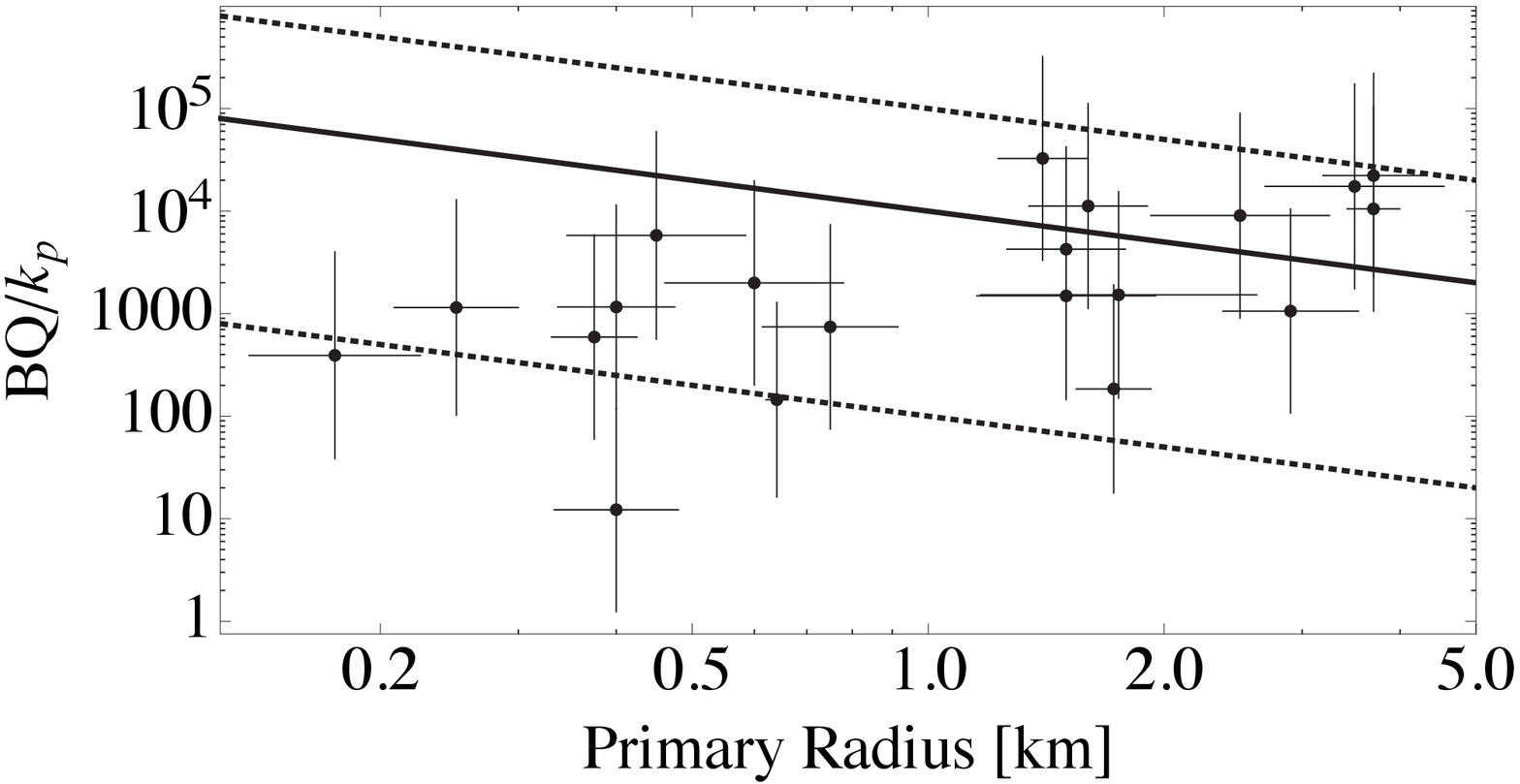}
\includegraphics[width=0.7\columnwidth]{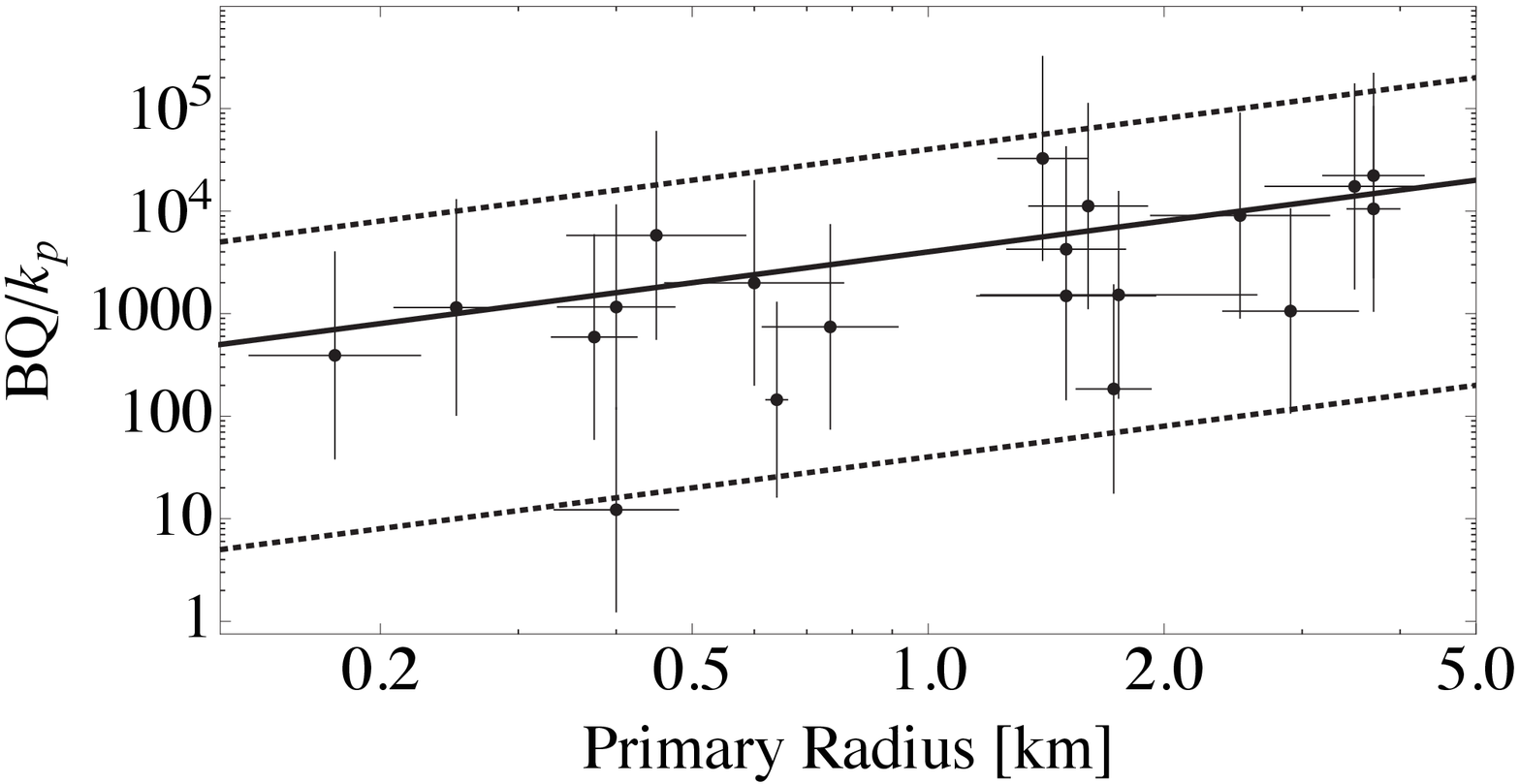}
\end{center}
\caption{$B Q / k_p$ were calculated directly from observed quantities according to equation~\ref{eqn:BQK} for each known synchronous binary, and plotted as a function of primary radius $R_p$ along with 1-sigma uncertainties (for clarity, the same data are shown in both plots). This data is also listed in Table 1. The solid line in the top plot shows the tidal Love number model from~\citet{Goldreich2009} for asteroids with ``rubble pile'' internal structures: $k_p = 10^{-5} R_p$. The bottom plot is a fitted model to the data: $k_p = 2.5 \times 10^{-5} R_p^{-1}$.  For both models, the tidal dissipation number $Q = 10^2$ and the BYORP coefficient $B = 10^{-3}$ is assumed. The dashed lines indicate the range of predicted scatter in the model due to the BYORP coefficient (possibly 10 times stronger or 100 times weaker).}
\label{fig:BQkPlot}
\end{figure}

If the observed synchronous population is assumed to be in this joint opposing evolutionary equilibrium state, then~$A = 1$ and Eqn.~\ref{eqn:semimajoraxisevolution} is solved for the three unknown quantities ($B$,~$Q$, and~$k_p$): 
\begin{equation}
\frac{B Q}{k_p} = \frac{2  \pi \omega_d^2 \rho R_p^2 q^{4/3}}{H_\odot a^7}
\label{eqn:BQK}
\end{equation}
Table~\ref{tab:BQk} lists and Fig.~\ref{fig:BQkPlot} plots~$B Q / k_p$ for each of the known synchronous binary systems using observational data~\citep{Pravec2006, Pravec2007b}. The tidal dissipation number $Q$ is an intensive property that we expect to be similar for all of these bodies, and for small bodies has been estimated to be $Q = 10^2$~\citep{Goldreich1966}. The scatter and size dependence in Fig.~\ref{fig:BQkPlot} should be from $B$ and $k_p$, respectively. The BYORP coefficient $B$ does not depend on size;~$B$ does depend on the shape of the secondary. Asteroid shapes can vary greatly introducing scatter in the BYORP coefficient.~\citet{McMahon2010b} estimate $B = 10^{-3}$ from the shape model of the secondary of 1999 KW$_4$, which does not appear symmetric. The BYORP coefficient may vary over a few orders of magnitude, especially towards smaller values corresponding to secondaries that are more symmetric. The tidal Love number may have a dependance on size, and~\citet{Goldreich2009} predict that the tidal Love number $k_p = 10^{-5} R_p$ for a ``rubble pile'' internal structure. 

In the top plot of Fig.~\ref{fig:BQkPlot}, the solid line plots a simple theoretical model of~$B Q / k_p$ using the estimates of each value from above, so that $BQ/k_p = 10^4 R_p^{-1}$. Fitting the proportionality constant of the~\citet{Goldreich2009} tidal Love number model does not significantly change the results. Dashed lines indicate scatter from the BYORP coefficient (one order of magnitude larger and two orders of magnitude smaller). This model works acceptably well for systems with primary radii $R_p = 2$ km but predicts values too large for systems with much smaller primaries. Deviations away from the model appear to be correlated with primary size, and so this may indicate that the tidal Love number has the wrong primary radius dependance. 

If a power law is fit to the logarithmic data, then the best fit is $BQ/k_p = 4 \times 10^3 R_p$ (shown as the solid line in the bottom plot of Fig.~\ref{fig:BQkPlot}). Using the same models for $B$ and $Q$ as above, then the tidal Love number dependence is $k_p = 2.5 \times 10^{-5} R_p^{-1}$. Systems with this tidal Love number dependence are stable in eccentricity at the long-term equilibrium in semi-major axis. Deviations from this model do not appear to have a dependance on primary radius, and furthermore the scatter follows our expectations for scatter due to the BYORP coefficient, namely that the largest deviations are about two orders below, while most systems fall within an order below and above.
\section{Conclusions}
\label{sec:Conclusions}
Small, synchronous binary asteroids may be residing in a long-term stable equilibrium. To inhabit the equilibria, asteroids cannot have a ``monolith'' internal structure but must have ``rubble pile'' interior. The tidal Love number is expected to vary with size, however the data suggest~$k \propto R^{-1}$, which is different than the~$k_R \propto R$ predicted by the~\citet{Goldreich2009} ``rubble pile'' theory. This may highlight the difficulty of using a modified continuum theory to model ``rubble piles.'' This suggests that the tidal Love number theory is incomplete, and future work should examine this closely. 

Current campaigns to detect the BYORP effect may not achieve the results predicted by previous studies~\citep{Pravec2010b, McMahon2010c}. Tides may have stopped the synchronous population from evolving long ago. Existence of a long-lived equilibrium prevents rapid disruption due to the BYORP effect, and so binary formation at a furious rate is no longer required to maintain a steady-state synchronous binary population. It may take many years of observation in order to conclude that systems are in this equilibrium and are truly no longer evolving. 

In the future, if $B$ is determined through secondary shape modeling for each system, the geophysical parameters $Q/k$ could be assessed directly.

\begin{deluxetable}{rccccccc} 
\tablecolumns{8} 
\tablewidth{0pc} 
\tablecaption{Properties and calculated $BQ/k_p$ of synchronous binary asteroid systems.} 
\tablehead{ 
\colhead{Asteroid System} & \colhead{$a_\odot$ (AU)}   & \colhead{$e_\odot$}    & \colhead{$\rho$ (g/cc)} & 
\colhead{$q$}    & \colhead{$R_p$~(km) }   & \colhead{$a$ ($R_p$)}    & \colhead{$BQ/k_p$}}
\startdata 
(1338) Duponta & $ 2.264$ & $ 0.113$ & $ 2.0^{ 1.0}_{ 0.7}$ & $ 0.014^{ 0.031}_{ 0.010}$ & $ 3.70^{ 0.59}_{ 0.51}$ & $ 4.00^{ 0.60}_{ 0.52}$ & $22191^{197448}_{19949}$ \\
(2044) Wirt & $ 2.382$ & $ 0.341$ & $ 2.0^{ 1.0}_{ 0.7}$ & $ 0.016^{ 0.035}_{ 0.011}$ & $ 3.50^{ 1.05}_{ 0.81}$ & $ 4.20^{ 0.63}_{ 0.55}$ & $17398^{155910}_{15652}$ \\
(2131) Mayall & $ 1.887$ & $ 0.111$ & $ 2.0^{ 1.0}_{ 0.7}$ & $ 0.027^{ 0.065}_{ 0.019}$ & $ 3.70^{ 0.30}_{ 0.27}$ & $ 4.80^{ 0.72}_{ 0.63}$ & $10506^{94037}_{9450}$ \\
(3309) Brorfelde & $ 1.818$ & $ 0.053$ & $ 2.0^{ 1.0}_{ 0.7}$ & $ 0.018^{ 0.039}_{ 0.012}$ & $ 2.50^{ 0.75}_{ 0.58}$ & $ 4.00^{ 0.60}_{ 0.52}$ & $9043^{80984}_{8134}$ \\
(5477) 1989 UH$_2$ & $ 1.917$ & $ 0.076$ & $ 2.0^{ 1.0}_{ 0.7}$ & $ 0.064^{ 0.147}_{ 0.045}$ & $ 1.50^{ 0.28}_{ 0.24}$ & $ 5.00^{ 0.75}_{ 0.65}$ & $4246^{37966}_{3819}$ \\
(6084) Bascom & $ 2.313$ & $ 0.236$ & $ 2.0^{ 1.0}_{ 0.7}$ & $ 0.051^{ 0.110}_{ 0.035}$ & $ 2.90^{ 0.64}_{ 0.52}$ & $ 7.40^{ 1.11}_{ 0.97}$ & $1060^{9407}_{953}$ \\
(7088) Ishtar & $ 1.981$ & $ 0.390$ & $ 2.0^{ 1.0}_{ 0.7}$ & $ 0.074^{ 0.159}_{ 0.051}$ & $ 0.60^{ 0.18}_{ 0.14}$ & $ 4.40^{ 0.66}_{ 0.57}$ & $1993^{17749}_{1792}$ \\
(9069) Hovland & $ 1.913$ & $ 0.118$ & $ 2.0^{ 1.0}_{ 0.7}$ & $ 0.064^{ 0.176}_{ 0.047}$ & $ 1.50^{ 0.45}_{ 0.35}$ & $ 5.80^{ 0.87}_{ 0.76}$ & $1490^{13818}_{1345}$ \\
(17260) 2000 JQ$_{58}$ & $ 2.204$ & $ 0.183$ & $ 2.0^{ 1.0}_{ 0.7}$ & $ 0.018^{ 0.041}_{ 0.012}$ & $ 1.60^{ 0.30}_{ 0.26}$ & $ 3.60^{ 0.54}_{ 0.47}$ & $11204^{100487}_{10080}$ \\
(31345) 1998 PG & $ 2.016$ & $ 0.391$ & $ 2.0^{ 1.0}_{ 0.7}$ & $ 0.064^{ 0.176}_{ 0.047}$ & $ 0.45^{ 0.14}_{ 0.10}$ & $ 3.40^{ 0.51}_{ 0.44}$ & $5802^{53808}_{5237}$ \\
(65803) Didymos & $ 1.644$ & $ 0.384$ & $ 2.0^{ 1.0}_{ 0.7}$ & $ 0.011^{ 0.024}_{ 0.007}$ & $ 0.38^{ 0.05}_{ 0.04}$ & $ 3.00^{ 0.45}_{ 0.39}$ & $591^{5258}_{531}$ \\
(66063) 1998 RO$_1$ & $ 0.991$ & $ 0.720$ & $ 2.0^{ 1.0}_{ 0.7}$ & $ 0.111^{ 0.242}_{ 0.076}$ & $ 0.40^{ 0.08}_{ 0.06}$ & $ 3.60^{ 0.54}_{ 0.47}$ & $1160^{10296}_{1043}$ \\
(66391) 1999 KW$_4$ & $ 0.642$ & $ 0.688$ & $ 2.0^{ 0.2}_{ 0.2}$ & $ 0.036^{ 0.077}_{ 0.025}$ & $ 0.64^{ 0.02}_{ 0.02}$ & $ 3.98^{ 0.12}_{ 0.12}$ & $145^{1139}_{128}$ \\
(76818) 2000 RG$_{79}$ & $ 1.930$ & $ 0.096$ & $ 2.0^{ 1.0}_{ 0.7}$ & $ 0.043^{ 0.093}_{ 0.029}$ & $ 1.40^{ 0.20}_{ 0.17}$ & $ 3.40^{ 0.51}_{ 0.44}$ & $32633^{288568}_{29318}$ \\
(85938) 1999 DJ$_4$ & $ 1.852$ & $ 0.483$ & $ 2.0^{ 1.0}_{ 0.7}$ & $ 0.125^{ 0.325}_{ 0.090}$ & $ 0.18^{ 0.05}_{ 0.04}$ & $ 4.20^{ 0.63}_{ 0.55}$ & $392^{3594}_{353}$ \\
(137170) 1999 HF$_1$ & $ 0.819$ & $ 0.463$ & $ 2.0^{ 1.0}_{ 0.7}$ & $ 0.012^{ 0.029}_{ 0.009}$ & $ 1.75^{ 0.88}_{ 0.58}$ & $ 3.40^{ 0.51}_{ 0.44}$ & $1525^{13970}_{1375}$ \\
(175706) 1996 FG$_3$ & $ 1.054$ & $ 0.350$ & $ 2.0^{ 1.0}_{ 0.7}$ & $ 0.030^{ 0.065}_{ 0.020}$ & $ 0.75^{ 0.17}_{ 0.14}$ & $ 3.80^{ 0.57}_{ 0.50}$ & $743^{6604}_{668}$ \\
(185851) 2000 DP$_{107}$ & $ 1.366$ & $ 0.377$ & $ 2.0^{ 1.0}_{ 0.7}$ & $ 0.069^{ 0.148}_{ 0.047}$ & $ 0.40^{ 0.08}_{ 0.07}$ & $ 7.20^{ 1.08}_{ 0.94}$ & $12^{108}_{11}$ \\
2002 CE$_{26}$ & $ 2.234$ & $ 0.559$ & $ 0.8^{ 0.3}_{ 0.2}$ & $ 0.001^{ 0.002}_{ 0.001}$ & $ 1.72^{ 0.20}_{ 0.18}$ & $ 2.72^{ 0.35}_{ 0.31}$ & $184^{1721}_{167}$ \\
2005 NB$_7$ & $ 2.044$ & $ 0.518$ & $ 2.0^{ 1.0}_{ 0.7}$ & $ 0.064^{ 0.262}_{ 0.051}$ & $ 0.25^{ 0.05}_{ 0.04}$ & $ 3.60^{ 0.54}_{ 0.47}$ & $1147^{11719}_{1045}$ \\
\enddata 
\tablecomments{Data  with 1-sigma uncertainties taken from the April 8, 2010 binary asteroid parameter release from \url{http://www.asu.cas.cz/~asteroid/binastdata.htm} as compiled by methods and assumptions described in~\citet{Pravec2006} and~\citet{Pravec2007b}.}
\label{tab:BQk}
\end{deluxetable}

\end{document}